\documentclass[twocolumn,aps,prl,showpacs,amsmath,superscriptaddress]{revtex4}
\usepackage{amssymb}
\usepackage{mathrsfs}
\usepackage{graphicx}
\usepackage{float}

\begin{document}
\title{Edge states and topological phases in one-dimensional optical superlattices }
\author{Li-Jun Lang}
\affiliation{Beijing National Laboratory for Condensed Matter
Physics, Institute of Physics, Chinese Academy of Sciences,
Beijing 100190, China}
\author{Xiaoming Cai}
\affiliation{Beijing National Laboratory for Condensed Matter Physics, Institute of
Physics, Chinese Academy of Sciences, Beijing 100190, China}
\author{Shu Chen}
\email{schen@aphy.iphy.ac.cn} \affiliation{Beijing National Laboratory for Condensed
Matter Physics, Institute of Physics, Chinese Academy of Sciences, Beijing 100190,
China}

\date{ \today}

\begin{abstract}
We show that one-dimensional quasi-periodic optical lattice
systems can exhibit edge states and topological phases which are
generally believed to appear in two-dimensional systems. When the
Fermi energy lies in gaps, the Fermi system on the optical
superlattice is a topological insulator characterized by a nonzero
topological invariant. The topological nature can be revealed by
observing the density profile of a trapped fermion system, which
displays plateaus with their positions uniquely determined by the
ration of wavelengths of the bichromatic optical lattice. The
butterfly-like spectrum of the superlattice system can be also
determined from the finite-temperature density profiles of the
trapped fermion system. This finding opens an alternative avenue
to study the topological phases and Hofstadter-like spectrum in
one-dimensional optical lattices.
\end{abstract}

\maketitle

{\it Introduction.-} In recent years ultracold atomic systems have
proven to be powerful quantum simulators for investigating various
interesting physical problems including many-body physics
\cite{Bloch} and topological insulators \cite{Kane}. In comparison
with traditional condensed matter systems, cold-atom systems
provide more control in constructing specific optical lattice
Hamiltonians by allowing both tunable hopping and trapping
potential to be adjusted as needed, thus optical lattices
populated with cold atoms offer a very promising alternative
avenue to build topological insulating states \cite{Kane}. Several
schemes exploring topological insulators with or without Landau
levels in optical lattices have been proposed
\cite{Zhu,Zhai,DasSarma,Spielman}. So far all these schemes focus
on two-dimensional (2D) systems, as one-dimensional (1D) systems
without additional symmetries are generally thought as lack of
topological nontrivial phases \cite{Ryu}.

In this work, we study properties of trapped fermions on 1D
quasi-periodic optical lattices and show that these systems
display non-trivial topological properties, which share the same
physical origins of topological phases of 2D quantum Hall effects
on periodic lattices \cite{Hofstadter}. The quasi-periodic optical
lattices can be generated by superimposing two 1D optical lattices
with commensurate or incommensurate wavelengths
\cite{Fallani,Roati,Deissler}, which has led to experimental
observation of Anderson localization of a noninteracting BEC of
$^{39}$K atoms in 1D incommensurate optical lattices \cite{Roati}.
Motivated by the experimental progress, one-dimensional optical
superlattices have been theoretically studied
\cite{Roscilde,Giamarchi,Yamashita,Buonsante,Chen}. For a
quasi-periodic superlattice, the single-particle spectrum is
organized in bands. When fermions is loaded in the superlattice,
the system shall form insulators if the chemical potential (Fermi
energy) lies in the gaps. For the open boundary system, localized
edge states are found to appear in the gap regimes. The appearance
of edge states is a signature indicating that the bulk states are
topological insulators characterized by nonzero Chern number. We
show that the topological invariant Chern number can be detected
from the density distribution of a trapped fermion system, which
displays plateaus in the local average density profiles with
positions of plateaus uniquely determined by the ration of
wavelengths of the bichromatic optical lattice. Through the
analysis of universal scaling behaviors in quantum critical
regimes of conductor-to-insulator transitions, we further display
that both the positions and widths of plateaus can be read out
from the finite-temperature density distributions, which provides
us an alternative way to study topological phases and
Hofstadter-like spectrum by using 1D optical superlattices. We
notice that localized boundary states in 1D incommensurate
lattices have been experimentally observed very recently by using
photonic quasi-crystals \cite{Kraus}.

{\it Quasi-periodic lattices.-} We consider a 1D polarized Fermi
gas loaded in a bichromatic optical lattice \cite{Roati,Deissler},
which is described by
\begin{equation}
H=-t\sum_i (\hat{c}^\dagger_i \hat{c}_{i+1}+
\mathrm{H.c.})+\sum_{i=1}^{L} V_i \hat{n}_i \label{H}
\end{equation}
with
\begin{equation}
\label{eqn2} V_i=V \mathrm{cos}(2\pi \alpha i+\delta),
\end{equation}
where $L$ is the number of the lattice sites, $\hat{c}^\dagger_i$
($\hat{c}_i$) is the creation (annihilation) operator of the
fermion, and $\hat{n}_i=\hat{c}^\dagger_i \hat{c}_i$. The hopping
amplitude $t$ is set to be the unit of the energy $(t=1)$, and $V$
is the strength of commensurate (incommensurate) potential with
$\alpha$ being a rational (irrational) number and $\delta$ an
arbitrary phase whose effect shall be illustrated lately. Suppose
that the n-th eigen-state of a single particle in the 1D lattice
is given by $|\psi_n \rangle=\sum_{i} u_{i,n} c_{i}^{\dagger}
|0\rangle$,  the eigenvalue equation $H|\psi_n\rangle=E_n
|\psi_n\rangle$ leads to the following Harper equation:
\begin{equation}
 -(u_{i+1,n}+u_{i-1,n})+V \mathrm{cos}(2\pi \alpha
i+\delta) u_{i,n}=E_n u_{i,n}, \label{Harper1}
\end{equation}
where $u_{i,n}$ is the amplitude of the particle wave function at
$i$-th site with $V_i$ the on-site diagonal potential and $E_n$
the $n$-th single particle eigenenergy. The ground state wave
function of the $N$ spinless free fermionic system can be written
as $|\Psi^G_F\rangle=\prod^N_{n=1}\sum^L_{i=1} u_{i,n}
c^\dagger_i|0\rangle $, where $N$ is the number of fermions.
\begin{figure}[tbp]
\includegraphics[scale=0.5]{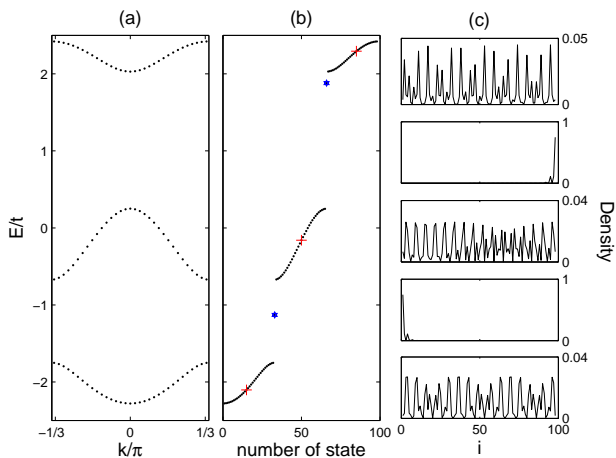}
\caption{(Color online) Spectrum of Hamiltonian (1) with
$\alpha=1/3$, $V=1.5$ and $\delta=2\pi/3$. (a) Energy bands for
the system with $99$ sites under periodic boundary conditions.
(b)Eigen-energies in ascending order for the system with $98$
sites under open boundary conditions. (c) From up to down, figures
represent the wavefunctions of states marked by labels (``+" for
states in bands and ``$\ast$" for states in gaps) in (b).}
\label{Fig1}
\end{figure}

The solution to Eq. (\ref{Harper1}) is closely related to the
structure of the potential $V_i$.  For the incommensurate case
$V_i=V \mathrm{cos}(2\pi\alpha i)$ with irrational $\alpha$,
Eq.(\ref{Harper1}) is the well-known Aubry-Andr\'{e} model
\cite{Aubry}, which has been showed that when $V<2$ all the single
particle states are extended and when $V>2$ all the single
particle states are the localized states. However, for a periodic
$V_i$, all the single particle states are extended band states
according to Bloch's theorem. Next we shall consider the
commensurate potential $V_i$ with a rational $\alpha$ given by
$\alpha=p/q$ with $p$ and $q$ being integers which are prime to
each other. Since the potential $V_i$ is periodic with a period
$q$, the wave functions take the Bloch form, which fulfills
$u_{i+q} = e^{ikq} u_{i} $, for the lattice under the periodic
boundary condition. Taking $ u_{j}=e^{ikj}\phi_{j}(k) $ for $|k|
\leq \pi/q$, we have $\phi_{j+q}(k)=\phi_{j}(k)$. In terms of
$\phi_{j}(k)$, Eq.(\ref{Harper1}) becomes
\begin{eqnarray}
& & -[e^{ik}\phi_{j+1} + e^{-ik}\phi_{j-1}]+V \mathrm{cos}(2\pi
 j p/q
+\delta) \phi_{j}\nonumber\\
 & &= E(k) \phi_{j}, \label{Harper2}
\end{eqnarray}
Since $\phi_{j+q}=\phi_{j}$, the problem of solving the Harper
equation Eq.(\ref{Harper2}) reduces to solving the eigenvalue
equation, $M \Phi =E \Phi $, where $\Phi=(\phi_1,\cdots,\phi_q)^T$
and $M$ is a $q \times q$ matrix. Solving the eigenvalue problem,
we get $q$ eigenvalues: $E_{\alpha}(k)$ with $\alpha=1,\cdots,q$.
Consequently, the energy spectrum consists of q bands. As an
example, the energy spectrum for a commensurate lattice with
$\alpha=1/3$ and $L=99$ under periodic boundary condition is given
in Fig.1a. For the lattice with open boundary condition, the
momentum $k$ is no longer a good quantum number. There appear edge
states in gap regimes as indicated by symbols of ``star" located
at gap regimes in Fig.1b. As shown in Fig.1c, these edge states
are localized in the left and right boundaries in contrast to
extended states marked by symbols of ``+" in band regimes.

As the phase $\delta$ varies from $0$ to $2\pi$, the spectrum for
a given $\alpha=p/q$ changes periodically. The position of the
edge states in the gaps also varies continuously with the change
of $\delta$. In Fig.2 we show the spectrum of the quasi-periodic
systems with $\alpha=1/3$ and $1/4$ versus $\delta$ under the open
boundary condition. The shade regimes correspond to the band
regimes and the lines between bands are the spectra of edge
states.

{\it Topological invariant.-} Appearance of edge states is
generally attributed to the nontrivial topological properties of
bulk systems  \cite{Kane,Hatsugai}.
Next we explore the topological properties of states under the
periodic boundary condition. The topological property of the
system can be understood in terms of Berry phase in $k$ space,
which is defined as $ \gamma = \oint A_k dk$, where $A_k$ is the
Berry connection $A_k = i \langle \phi(k)|
\partial_k |\phi(k) \rangle $ and $\phi(k)$ the occupied Bloch
state \cite{Xiao}. Adiabatically varying the phase $\delta$ from
$0$ to $2\pi$, we get a manifold of Hamiltonian $H(\delta)$ in the
space of parameter $\delta$. Similarly, we can define the Berry
connection $A_\delta = i \langle \phi(k,\delta)|
\partial_\delta |\phi(k,\delta) \rangle $. For eigenstates $\phi(k,\delta)$ of $H(\delta)$, we
may use the Chern number to characterize their topological
properties. The Chern number is a topological invariant which can
be calculated via $C= \frac{1}{2 \pi } \int_{0}^{{2\pi}/{q}}{dk}
\int_{0}^{2\pi} {d \delta} [\partial_k A_\delta -
\partial_\delta A_k ]$.
To calculate it, we need work on a discretized Brillouin zone.
Here we follow the method in Ref. \cite{Fukui} to directly perform
the lattice computation of the Chern number. For the system with
$\alpha=1/3$, we find that the Chern number for fermions in the
lowest filled band (with $1/3$ filling) is $1$, while the Chern
number for states with the second band fully filled (with $2/3$
filling) is $-1$.

To understand the topological origin of fermions in the 1D
quasi-periodic lattices, we explore connection of the present
model to the well-known Hofstadter problem
\cite{Harper,Hofstadter}, which describes electrons hopping on a
2D square lattice in a perpendicular magnetic field, with the
Hamiltonian given by $H=-\sum_{\langle i,j \rangle} t_{ij}
\hat{c}^\dagger_j \hat{c}_{i} e^{i2\pi \phi_{ij}}$, where the
summation is over nearest-neighbor sites and the magnetic flux
through each plaquette given by $\phi=\sum_{plaquette} \phi_{ij}$.
Taking the Landau gauge, the eigenvalue problem is described by
the Harper equation \cite{Hofstadter}: $-t_x
(\psi_{j-1}+\psi_{j+1})-2t_y \cos( 2 \pi j \phi- k_y ) \psi_j=
E(k_y) \psi_j$, where $t_x$ ($t_y$) is the hopping amplitude along
the $x$ ($y$) direction. If we make substitutions of $t
\rightarrow t_x$, $V \rightarrow -2t_y$, $\alpha \rightarrow \phi$
and $\delta \rightarrow -k_y$, the current 1D problem can be
mapped to the lattice version of the 2D integer Hall effect
problem. For the latter case, it was shown that when the chemical
potential lies in gap regimes, the Hall conductance of the system
is quantized \cite{TKNN} and the corresponding Hofstadter
insulating phase is a topological insulator characterized by
nonzero Chern number.

Keeping this connection in mind, it is not strange to see that the
energy spectrum of the 1D quasi-periodic system versus different
$\alpha$ has similar structure as the spectrum of the 2D
Hofstadter butterfly. In Fig.{\ref{Fig3}}a and b, we show the
spectrum for the system with phase $\delta=0$ and $\delta=\pi/4$,
respectively. The basic structure is quite similar for different
phases except for some minor differences, for example, the
spectrum for $\alpha=\pi/4$ is continuous around $E=0$ for
$\delta=0$, but separated for $\delta=\pi/4$ (see also Fig2.b).
The familiar Hofstadter spectrum is actually a summation over all
the 1D spectra with different phases $\delta$ from $0$ to $2 \pi$.
Despite the existence of the mapping between the 1D quasi-periodic
system and the 2D Hofstadter system, we note that the edge modes
in the 1D system are localized and would not be used for
dissipationless transport as in high-dimensional systems.

\begin{figure}[tbp]
\includegraphics[scale=0.46]{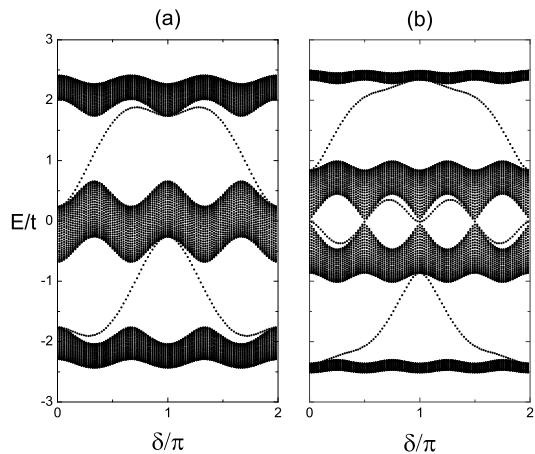}
\caption{ Energies varying with the phase $\delta$ for systems
with $V=1.5$, (a) $\alpha=1/3$, $L=98$, and (b) $\alpha=1/4$,
$L=99$ under open boundary conditions.} \label{Fig2}
\end{figure}
\begin{figure}[tbp]
\includegraphics[scale=0.46]{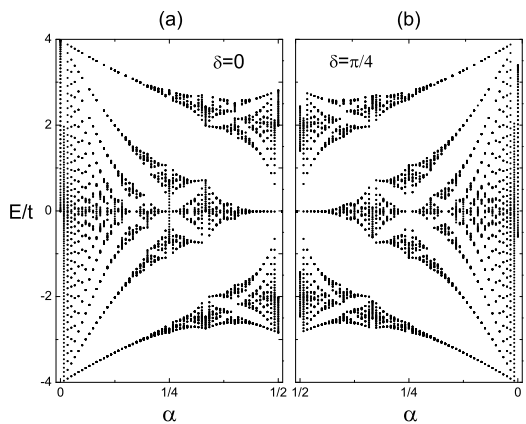}
\caption{ The butterfly-like energy spectra with respect to
$\alpha$ varying from $0$ to $1/2$ with different phases: (a)
$\delta=0$, (b) $\delta=\pi/4$. Both figures are for the system
with $V=2$ and $L=120$ under periodic boundary conditions.}
\label{Fig3}
\end{figure}

{\it Experimental detection.-} In realistic ultracold atom
experiments, we need consider the effect of an external confining
potential, i.e., $V_i$ in Eq.(\ref{H}) is given by
\begin{equation}
V_i=V \mathrm{cos}(\alpha2\pi i+\delta)+V_H(i-i_0)^2, \label{V2}
\end{equation}
where $V_H$ is the strength of the additional harmonic trap with
$i_0$ being the position of trap center. The density distribution
for the trapped system can be calculated via
$n_i=\langle\Psi^G_{\mathrm{F}}| \hat{n}_i
|\Psi^G_{\mathrm{F}}\rangle$ with $\Psi^G_{\mathrm{F}}$ the ground
state wavefuction. In Fig.4, we have shown the local average
density distribution of fermions trapped in both commensurate and
incommensurate optical lattices with a harmonic trap. Here, in
order to reduce the oscillations in density profiles induced by
the modulation of potentials, we have defined the local average
density $\bar{n}_i=\sum_{j=-M}^M n_{i+j}/(2M+1)$, where $2M+1$ is
the length to count the local average density with $M\ll L$
\cite{Cai}. After counting on the locations of plateaus, we find
that heights of plateaus are completely decided by $\alpha$ with
values $\bar{n}(\alpha)=\alpha, 1-\alpha, 2\alpha, 1-2\alpha,
4\alpha, 1-4\alpha,...$, if the values are in the range of
$(0,1)$. The plateaus with $\bar{n}_i=\alpha, 1-\alpha$ are the
widest ones corresponding to the largest gap regimes in the
butterfly spectrum, while $\bar{n}_i=2\alpha, 1-2\alpha$
correspond to the smaller gap regimes. The width of plateau is
associated with the size of gap.

One can also understand the appearance of plateaus under the
local-density approximation (LDA), in which a local chemical
potential is defined as $\mu(i)=\mu-V_H(i)$, where $V_H(i)$ is the
harmonic trap potential and $\mu$ is determined by $\sum_i
n(\mu(i),T) = N$. The LDA is applicable provided that the number
of fermions is large, and the variation of the trap potential is
slow. Within the LDA, the local chemical potential $\mu(i)$
decreases parabolically away from the center of the trap. When the
local chemical potential lies in one of the gaps, there appear
plateaus in the density profile. The discernible number of
plateaus is related to the size of the energy gaps. For instance,
in Fig. 4, the plateau with $n=1$ corresponds to the band
insulator with completely filled band, which is topologically
trivial with zero Chern number. For $\alpha=1/3$ the chemical
potential passes through two gap regions which produces two
plateaus with $n=1/3$ and $n=2/3$, whereas for $\alpha=1/5$ the
chemical potential passes through four gap regions which gives
four plateaus with $n=1/5,2/5,3/5,4/5$, respectively. The Chern
number can be obtained from the density by using the Streda
formula \cite{Streda,Zhai}, which is valid when the chemical
potential lies in a gap. From the Streda formula $C=\frac{\partial
\bar{n} (\alpha)}{\partial \alpha}$, we can get $C=1,-1$ for
$\bar{n}_i=\alpha, 1-\alpha$ and $C=2,-2$ for $\bar{n}_i=2\alpha,
1-2\alpha$.
\begin{figure}[tbp]
\includegraphics[scale=0.55]{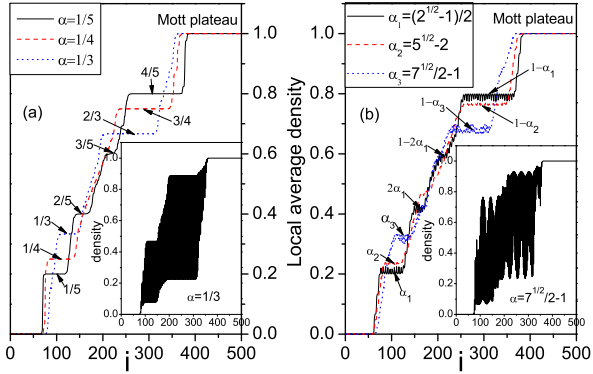}
\caption{ The local average density profiles for systems with
rational $\alpha$ (a) and irrational $\alpha$ (b). Inserts for
both pictures are the corresponding density profiles. The system
is with 1000 sites, 600 free fermions, $V=1.5, \delta=0,
V_H=3\times 10^{-5}$. Here we take $M=q$ for the rational case of
(a) and $M=20$ for the irrational case of (b).}. \label{Fig4}
\end{figure}
\begin{figure}[tbp]
\includegraphics[scale=0.55]{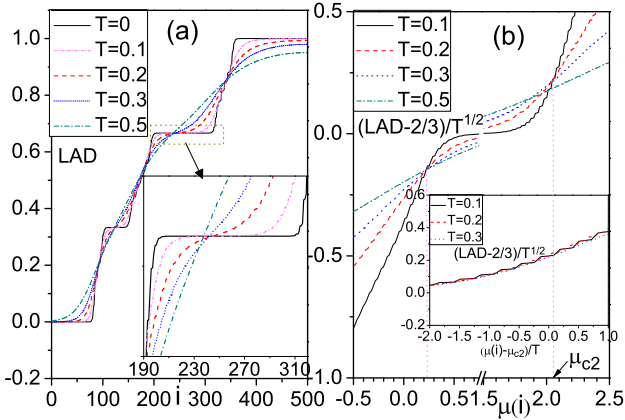}
\caption{ (a): Local average density profiles for systems with
different temperatures. Insert: enlargement of the plateau area.
(b): Scaled local average density profles vs $\mu(i)$ at different
temperatures. Insert: the universal function around critical point
$\mu_{c_2}$ for the corresponding system. The system is with 1000
sites, 600 free fermions , $V = 1.5, \alpha= 1/3,   \delta=0,$ and
$V_H = 3 \times 10^{-5}$. }. \label{Fig5}
\end{figure}

In principle, all gaps in the spectrum, including the nontrivial
smaller gaps in the butterfly, can be determined via observing the
corresponding plateaus. Widths of plateaus are proportional to
sizes of corresponding energy gaps. Experimentally it becomes
increasingly harder to observe these smaller gaps, as the
corresponding plateaus are very narrow which needs more precisely
experimental measurement of the density distribution. For the
realistic detection, we need consider the effect of temperature.
The finite-temperature density distribution can be calculated via
$ n_i(T)=\frac{1}{Z}\sum_{n=1}^{N_s} e^{-E_n/k_B T}
\langle\Psi^n_{\mathrm{F}}| \hat{c}^\dagger_i
\hat{c}_i|\Psi^n_{\mathrm{F}}\rangle$, where $N_s=L!/(L-N)!N!$ is
the number of states, $E_n$ is the energy of eigenstate
$\Psi^n_{\mathrm{F}}$, and $Z=\sum_{n=1}^{N_s} e^{-E_n/k_B T}$ is
the canonical partition function \cite{Rigol}. In Fig.\ref{Fig5}a,
we display the local average density profiles at different
temperatures.  It is shown that plateaus already become invisible
for $T>0.3t$. Despite the fact that the obvious plateaus are
smeared out by temperature fluctuations, we demonstrate that the
position and the width of zero-temperature plateaus can be
uniquely mapped out from finite-temperature density distributions
of the trapped optical lattice system, which fulfil some universal
scaling relations \cite{Zhou} near the zero-temperature phase
transition point $\mu=\mu_c$:
\begin{equation}
\label{eqn3} n(\mu,T)-n_r(\mu,T)=T^{\tfrac{d}{z}+1-\tfrac{1}{\nu
z}}\Omega(\tfrac{\mu-\mu_c}{T^{1/\nu z}}),
\end{equation}
where $n=n(T,\mu)$ represents the density distribution for
fermions with temperature $T$ and chemical potential $\mu$,
$n_r(\mu,T)$ is the regular part of the density, $\Omega(x)$ is a
universal function describing the singular part of density near
criticality, $d=1$ the dimensionality of the system, $\nu$ the
correlation length exponent, and z the dynamical exponent. It was
shown that $z=2$ and $\nu=1/2$ for the metal-insulator transition
of 1D free fermions \cite{Zhou}. From Fig.5b, we see that the
different curves of $[\bar{n}(\mu,T)-2/3]/T^{3/2}$ intersect at
two points. From intersecting points, we can determine $\mu_{c_1}$
and $\mu_{c_2}$, and thus the corresponding gap in the spectrum
can be inferred from $\mu_{c_2} - \mu_{c_1}$. In terms of the
scaled chemical potential $\widetilde{\mu}=(\mu-\mu_c)/T$, curves
for different temperatures collapse into a single one as shown in
the inset of Fig.5b.

In order to connect to the real experiment in cold atoms, we refer
to the parameters in Ref. \cite{Kohl} in which $^{40}K$ atoms were
used in an optical lattice with spacing $a=413$ nm, the deep of
the potential $V_{0}=5E_{R}$, and recoil energy $E_{R}=45.98\hbar$
kHz, thus the hopping magnitude $t=0.066E_{R}=3.035\hbar$ kHz.
Correspondingly, the temperature $T$ in Fig. 5 in unit $t/k_{B}$,
where $k_{B}$ is the Boltzmann constant, is of the order of $10$
nK, e.g., $T=0.1$ corresponds to 2.3084 nK.

{\it Summary.-} In summary, we explored edge states and
topological nature of Fermi systems confined in 1D optical
superlattices. Our study reveals that the topological invariant
can be detected from plateaus of density profiles of the trapped
lattice systems. Our results also clarify the connection of 1D
superlattice systems to 2D Hofstadter systems and will be useful
for studying topological phases and observing the Hofstadter
spectrum by using 1D optical lattices.

\begin{acknowledgments}
SC would thank Prof. S. Q. Shen for helpful discussions on edge
states of the dimerized chain which stimulates our interest in the
1D topological insulator. While we are preparing the present
paper, we become aware of the experiment for the observation of
edge states in 1D quasi-crystals \cite{Kraus}. This work has been
supported by NSF of China under Grants No.10821403, No.11174360
and No.10974234, 973 grant and National Program for Basic Research
of MOST.
\end{acknowledgments}

\end{document}